\documentclass[apj]{emulateapj}

\usepackage{color}
\usepackage{soul}

\shorttitle{SDSSJ0159+0105: a SMBH Binary Candidate Inside?}
\shortauthors{Z.-Y. Zheng,  et al. 2015}

\begin{document}

\title
{
 SDSS J0159+0105: A Radio-Quiet Quasar with a Centi-Parsec Supermassive Black Hole Binary Candidate$^*$
}

\author{Zhen-Ya Zheng\altaffilmark{1, 2, 3$^{a,b}$}}
\author{Nathaniel R. Butler$^{2}$}
\author{Yue Shen$^{4, 5}$} 
\author{Linhua Jiang$^{6}$}
\author{Jun-Xian Wang$^{7}$}
\author{\\Xian Chen$^{1}$}
\author{Jorge Cuadra$^{1}$}

\affil{$^1$Instituto de Astrofisica, Pontificia Universidad Catolica de Chile, 7820436 Santiago, Chile; zzheng@astro.puc.cl \\
$^2$ School of Earth and Space Exploration, Arizona State University, Tempe, AZ 85287, USA\\
$^3$ Chinese Academy of Sciences South America Center for Astronomy, 7591245 Santiago, Chile \\
$^4$ Department of Astronomy, University of Illinois at Urbana-Champaign, Urbana, IL 61801, USA \\
$^5$ National Center for Supercomputing Applications, University of Illinois at Urbana-Champaign, Urbana, IL 61801, USA \\
$^6$ The Kavli Institute for Astronomy and Astrophysics, Peking University, Beijing, 100871, China \\
$^{7}$CAS Key laboratory for Research in Galaxies and Cosmology, Department of Astronomy, University of Science and Technology of China,\\\nonumber 
Hefei, Anhui 230026, China
}

\altaffiltext{*}{This paper uses data from Sloan Digital Sky Survey (SDSS), Catalina Real-time Transient Survey (CRTS), {\it SWIFT}, {\it GALEX}, {\it 2MASS}, and {\it WISE} archive.}
\altaffiltext{a}{China-Chile CAS-CONICYT Fellow.}
\altaffiltext{b}{present address:  Instituto de Astrofisica, 
Pontificia Universidad Catolica de Chile,  Vicuna Mackenna 4860, 7820436 Macul, Santiago, Chile}

\begin{abstract}

We report a candidate centi-parsec supermassive black hole binary (SMBHB) in the radio-quiet quasar SDSS\,J0159+0105 at $z=0.217$. 
With a modified lomb-scargle code ({\it GLSdeDRW}) and the auto-correlation analysis (ACF), we detect two significant (at $P>99\%$) periodic signals at $\sim$\,741 day and $\sim$\,1500 day
from the 8.1-year Catalina V-band light curve of this quasar. The period ratio, which is close to 1:2, is typical of a black-hole binary system with a mass ratio of 0.05$<q<$0.8 according to recent numerical simulations.
SDSS\,J0159+0105 has two SDSS spectroscopic observations separated by $\sim$\,10 years. There is a significant change in the broad H$\beta$ profile between the two epochs, which can be explained by a single broad-line region (BLR) around the binary system illuminated by the aforementioned mini-disks, or a stream of gas flowing from the circumbinary disk to one of the SMBHs.
From the single BLR assumption and the orbital period $t_{orb}\sim$\,1500 day, we estimate the total virial masses of M$_{\rm SMBHB}$ $\sim$ 1.3$\times$10$^8$M$_\odot$, the average distances of 
BLR of $\sim$0.04pc ($\sim$50 light-day, with $\pm$0.3 dex uncertainty), and a SMBHB separation of $d$ = (0.01pc)$M_{8,tot}^{1/3}$ (T$_{\rm rest}$/3.3yr)$^{2/3}$ $\sim$ 0.013 pc (15 light-day). Based on analytical work, the postulated circumbinary disk has an inner radius of 2$d$ = 0.026 pc (30 light-day). SDSS\,J0159+0105 also displays unusual spectral energy distribution.
The unique properties of SDSS\,J0159+0105 are consistent with it being a centi-parsec SMBHB.
\end{abstract}


\section{Introduction}
\label{sec:intro}

The formation of supermassive black hole binaries (SMBHBs) is an inevitable consequence of frequent galaxy mergers during the hierarchical formation of galaxies \citep[e.g.,][]{Begelman80}. However, a theoretical question remains open: Whether two supermassive black holes (SMBHs) in such a binary
can evolve into a separation below $1$ pc and eventually coalesce \citep[for a review on the ``final parsec'' problem, see, e.g.,][and references therein]{Vasiliev14}. Given the anticipation of detecting low-frequency gravitational waves from merging SMBHs by the ongoing and upcoming experiments \citep[e.g., PTA, eLISA,][]{Hobbs10,Amaro-Seoane13}, it is of critical importance to identify sub-parsec SMBHB targets with various conventional astronomical approaches. 

Several methods have been used earlier to search for sub-parsec SMBHBs. The spectroscopic monitoring method utilizes the sub-parsec sizes of quasar broad-line regions (BLRs), and looks for coherent acceleration of the broad-line centroid due to the orbital motion of the binary \citep[e.g.,][]{Gaskell83}. This method is sensitive to SMBHBs with sub-pc separations, where the orbital acceleration is large enough to be measured from spectroscopy separated by several years 
\citep[e.g.,][]{ShenLoeb10}. Numerous studies have practiced this method and reported sub-pc SMBHB candidates which showed broad emission lines with significant systematic velocity offsets in single-epoch spectroscopy \citep[e.g.,][]{Komossa08,BorosonLauer09,Shields09,Tsalmantza11} or velocity acceleration in multi-epoch spectroscopies \citep[e.g.,][]{Eracleous12,Ju13,Shen13,Liu14,Runnoe15}. 
It is worth noting that the binaries found with this method have a typical orbital period of hundreds of years.

A second method, which only recently becomes possible to apply to large samples, is to search for periodic variations in quasar light curves derived from long-term photometric monitoring. Limited by the time baselines of most photometric monitoring programs, this method is mostly applicable to smaller-separation SMBHBs with a period of a few years\footnote{One exception is the sub-pc SMBHB candidate OJ 287, discovered from a century-long light curve, which shows a pair of outbursts every 12.2 years \citep[e.g.,][]{Sillanpaa88,Valtonen08}.}. In particular, two candidates with sinusoidal light curves have been discovered recently in optical bands\footnote{Multi-year monitoring of radio-loud quasars have also reported periodic variations in the radio light curves, some of which bear remarkable resemblance to the optical light curves in PG 1302-102 \citep[e.g.,][]{ Kudryavtseva11}.}. \citet{Graham15a} reported a SMBHB candidate in the quasar PG\,1302-102 based on a 5.2-year period from a 9-year optical monitoring by the Catalina Real-time Transient Survey, and \citet{Liu15} reported another candidate in the quasar PSO\,J334.2028+01.4075 from the Pan-STARRS1 survey. However, these two SMBHB candidates are hosted by radio loud quasars. There are alternative models to the binary one for radio-loud quasars with periodicities, i.e., precession jets. Several groups \citep[i.e.,][and references therein]{Zhang14,Sandrinelli14,Sandrinelli16} had reported blazars showing periodicities both in the optical/NIR and gamma rays.

A third method relies on a theoretical modeling of the spectral energy distribution (SED) of the emission from SMBHBs. Recent theoretical work on SMBHB evolution and accretion suggested distinctive features in the emitting SED, such as a flux deficit in the UV/optical band due to the opening of a gap in the accretion disk by the tidal perturbation of the binary \citep{ArmitageNatarajan02}. 
  \citet{Yan15} reported such a UV/optical deficit in Mrk 231 as evidence of SMBHB. However, \citet{Leighly16} demonstrated that the SMBHB model in Mrk 231 is untenable.
In contrast, more recent theoretical work predicted a significantly larger (instead of lower) high frequency radiation \citep{Lodato09, Farris15}, which could be caused by the mini-disks and the shocked streams inside the cavity. Note that the brightening and dimming on the high frequency radiation can also be explained as AGN flare and transient obscuration, respectively.  Considering the diversity on the SED analysis, it is critical to give a discriminant hint about the presence of a SMBHB through SED fitting.

All these methods have their own caveats, and alternative interpretations (with single SMBHs) exist for the observational signatures mentioned above. To confirm the existence of SMBHBs in these candidates,  
ideally one would carry out follow-up observations to look for further, different evidence. 

In this work, we report a SMBHB candidate in the quasar SDSS\,J0159+0105 at $z=0.217$, 
initially identified using the photometric periodicity method,
and also supported by the unique properties in the overall SED from radio to X-ray as well 
as in multi-epoch optical spectroscopy. Unlike the other two SMBHB candidates identified 
with optical periodicity, SDSS\,J0159+0105 is a radio-quiet quasar, thus the interpretation 
of the periodic signal is less uncertain (i.e., the periodic signal cannot be produced by a 
jet). In Section 2, we describe a new, robust method to search for periodic signals in quasar 
light curves in which we take into account the stochastic variability of quasars. In Section 3, 
we summarize the observational properties of SDSS\,J0159+0105, including the photometric time 
series, the SED from radio to X-ray, and the peculiar spectroscopic features in the broad lines. 
In Section 4, we interpret the peculiar properties of SDSS\,J0159+0105 in the context of the 
binary model, and we derive physical parameters for the binary based on the current theoretical 
understanding of SMBHB systems. We briefly summarize our main conclusions in Sec. 5.

\section{Searching Periodic Signals in Quasar Light Curves}
\label{sec:vtest}

\begin{figure}
\begin{center}
\includegraphics[angle=0,width=\linewidth,width=\linewidth]{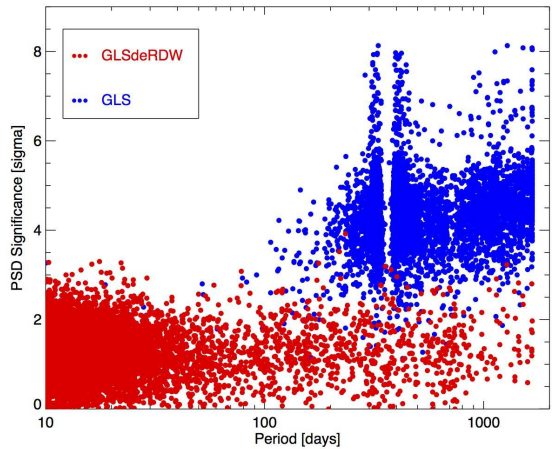}
\caption{This shows the results of running {\it GLS} and {\it GLSdeDRW} tests on the 6308 quasars SDSS Stripe 82 light curves from \citet{Ivezic07}. 
GLS algorithm will produce many false detections in quasar light curves, while the GLSdeDRW is more robust against false positives.
Here {\it GLSdeDRW}  pulls back the period and significance of the peaks so that they appear to be randomly distributed.  }
\label{qsoperiods}
\end{center}
\end{figure}

\begin{figure*}
\begin{center}
\includegraphics[angle=270,width=\linewidth]{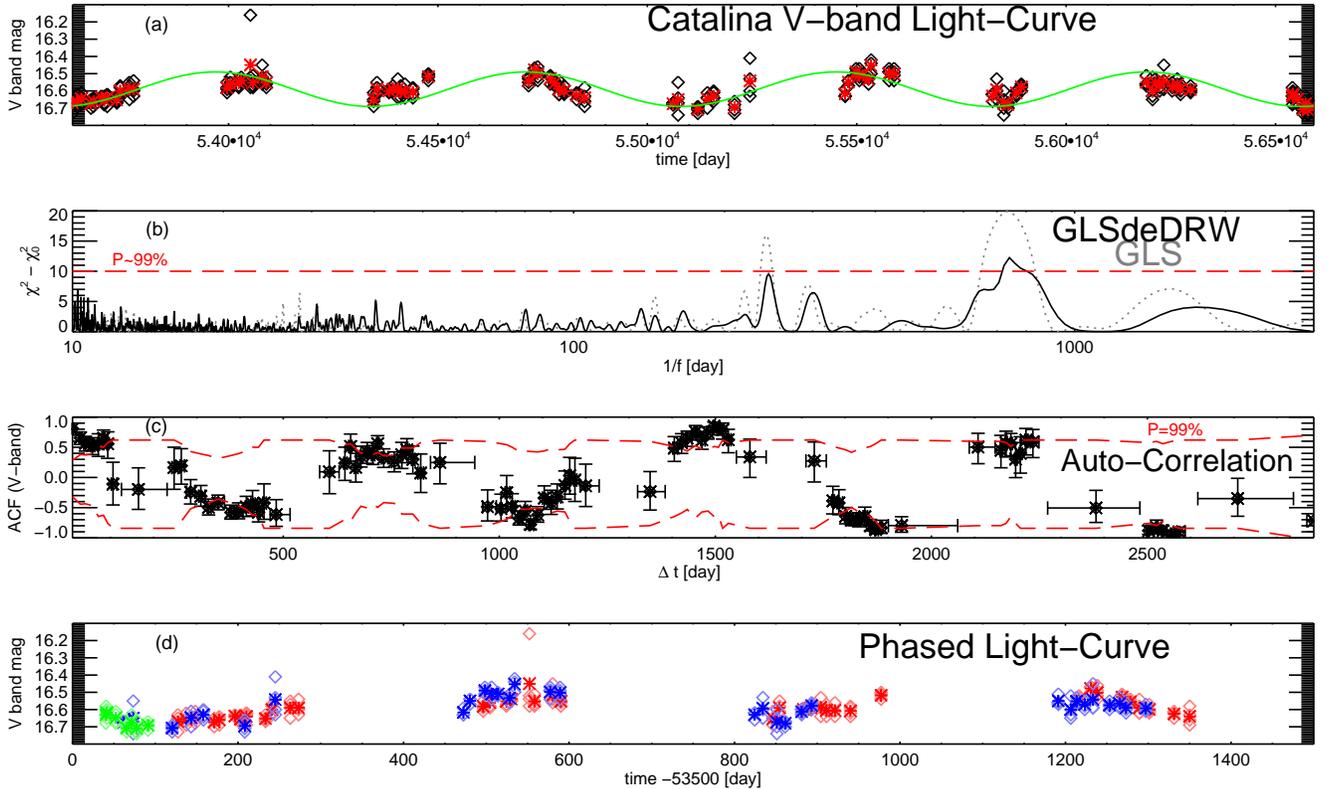}  
\caption{The light curves and time series analysis of SDSS J0159+0105. Panel (a): the raw (empty diamonds) and binned (1-day lag, 
red asterisks) Catalina $V$-band light curve of SDSS J0159+0105. Panel (b): The time series analysis with {\it GLS} (grey dotted curves) and 
{\it GLSdeDRW} (black curves). The red long-dashed line presents the $\sim$99\% significant limit of {\it GLSdeDRW} analysis. Panel (c): 
The auto-correlation analysis of the binned V-band light curve of  J0159+0105.  The long-dashed red line presents the 99\% significant 
level of the auto-correlation analysis. Panel (d): The Catalina $V$-band light curve of J0159+0105 in the phased T=1500 day light curve. The 
red, blue, and green symbols are light curves in the 1st, 2nd and 3rd phases, respectively. The empty diamonds and filled asterisks are raw and binned (1-day lag) light curve. 
}
\label{ppn129lc}
\end{center}
\end{figure*}

A common tool used to search for periodic signals in unevenly sampled time series is the Lomb-Scargle \citep[hereafter LS,][]{Lomb76,Scargle82} periodogram. This method can be regarded as a least-squares fit of sine and cosine functions to an observed time series. The LS
method has been generalized (hereafter GLS) to allow for measurement errors and also a mean offset in the observed time series 
\citep[e.g.,][]{ZechmeisterKuerster09,Richards11}, important for when phase sampling is non-uniform. 
However, when applied to quasar light curves, GLS tends to report many false detections of periods on the order of the duration of the light curve, as demonstrated in Fig. \ref{qsoperiods} and explained below.
This is because all quasars, unlike most stars for example, have light curves which meander slowly in time, yielding ``red'' power spectra with 
significant noise power on long timescales \citep[e.g.,][]{ButlerBloom11}. Traditional methods to evaluate LS period significance (or equivalently 
false alarm probability) assume white noise spectra; these fail and produce large over-estimates \citep[e.g.,][]{Vaughan04} of significance as well 
as many false alarms when the noise is red. To overcome this problem it is necessary to employ correct assumptions in evaluating the significance 
of periods when fitting sines and cosines to quasar light curves.

The limitations of a traditional LS period search can easily be overcome by generalizing the least-squares fitting process using the correct 
likelihood function for quasars. Recent work demonstrates that quasar light curves are well described as a damped random walk (DRW) 
or Ornstein-Uhlenbeck (OU) process \citep{Kelly09, Kozlowski10, Mushotzky11}. This stochastic approximation is 
especially powerful when separating quasars from stars through their optical variability \citep{ButlerBloom11, MacLeod11}. 
In the DRW model, the covariance matrix is:
\begin{equation}
    C_{ij} = \sigma^2_i \delta_{ij} + {1 \over 2} \hat \sigma^2 \tau_{\circ} \exp{( -\tau_{ij}/\tau_{\circ} )}.
    \label{eq:omega}
\end{equation}
In the traditional LS approach for white noise data, only the first term above is present, where the $\sigma_i$ are the measurement errors.
The second term implies covariance between epochs and leads to the red power spectra. \citet{ButlerBloom11} have demonstrated that 
the two parameters (variability magnitude scale $\hat \sigma$ and covariance time scale $\tau_{\circ}$) -- which fully define the DRW 
likelihood -- are set a priority by the observed quasar brightness alone.  Hence, the likelihood is defined prior to the period search.  
This allows for an efficient and statistically rigorous search for periodic signals, taking into account an accurate estimate of the 
true randomness expected for quasars of a given brightness. We have implemented software in python to extend the generalized LS method 
to this likelihood and applied this software {\it GLSdeDRW}\footnote{Please see the code at http://butler.lab.asu.edu/qso\_period/} here.

We have conducted blind search for periodic signals on the light curves of bright quasars in the SDSS ``Stripe 82'' field.\footnote{The equatorial Stripe 82 
region (20h24m\,$<$\,R.A.\,$<$\,04h08m, -1.27deg\,$<$\,Dec.\,$<$\,+1.27deg, Area $\sim$ 290 deg$^2$)
was repeatedly observed --- 58 SDSS imaging runs from 1998 September to 2007 December --- with 1--2 observations per week, each Fall.}
The software {\it GLS} and {\it GLSdeDRW} are applied to the light curves of  6308 SDSS Stripe 82 quasars from the variable source catalog of \citet{Ivezic07}. 
However, we didn't find any $>$4-$\sigma$ significant signals from all these quasar light curves with {\it GLSdeDRW} (see Fig. \ref{qsoperiods}).
It could be caused by the poor cadence of SDSS Stripe 82 light curves, as the numbers of observing epochs in the first few years are significantly less than that of the 
last three years. It could also be due to the rareness of the periodic signals in quasar light curves.

In addition to SDSS, this field is also covered by the Catalina Real-time Transient Survey \citep[CRTS,][]{Drake09, Djorgovski11, 
Mahabal11}. We download the CRTS $V$-band light curves from the Catalina archive server\footnote{Link: http://catalinadata.org}. The baseline 
length of CRTS ($\sim$ 9 years per source) is similar to that of SDSS Stripe 82 survey , but CRTS is a single $V$-band time series with 
$\sim$250 observations evenly distributed in each year, while SDSS Stripe 82 is a five-band ({\it ugriz}) time series with $\sim$70 observations in each band unevenly distributed in each year. 
From SDSS DR7 quasar catalog \citep{Shen11}, there are about 9000 quasars in Stripe 82 with SDSS spectroscopic confirmation, and $\sim$1500 quasars in the redshift range of 
0.15 $< z <$ 0.8. Here we focus on the low-$z$ quasar sample with good spectroscopic qualities (SDSS spectral S/N $>$ 10 per pixel), which leaves 347 quasars in the redshift range of 0.15 $< z <$ 0.8. 
In the following analysis we search for the periodic signals from CRTS light curves for these 347 bright low-$z$ quasars. 
We also require that the periodic signal is visible in the auto-correlation analysis (ACF) of CRTS light curve, and in the cross-correlation analysis (CCF)
between the CRTS and SDSS Stripe 82 light curves.  

For both the {\it GLS} and {\it GLSdeDRW} test, we choose a frequency range from 1/$T_{max}$ to $N_0$/(2$T_{max}$) day$^{-1}$ with a step size of 
($N_0$/2-1)/($T_{max}$$N_i$)$\sim$ 1/8000 day$^{-1}$. Here $T_{max}$ $\sim$ 3300 days is the baseline of the light curves, $N_0$ is the number 
of observed epochs, and $N_i$ is the number of independent frequencies from \citet{HorneBaliunas86}.  We follow the false alarm probability (FAP) calculations in 
\citet[][Eq. 24]{ZechmeisterKuerster09}, which are:
\begin{eqnarray}
FAP & = & 1 - [1-(1-p_{best})^{\frac{N_0-1}{2}}]^{N_i} \\
FAP & \approx & N_i\times(1-p_{best})^{\frac{N_0-1}{2}} \qquad \text{for   FAP } \ll 1,
\end{eqnarray}
here $p(f)$ = $\frac{\chi^2_0 - \chi^2(f)}{\chi^2_0}$ is the normalized power spectrum, which is about the difference in the goodness of fitting, $\chi^2_0$ is fitting a constant to the light curve, and $\chi^2$ is the value when fitting a constant as well as sine and cosine. The difference $\chi^2_0 - \chi^2(f)$ should be $\chi^2$ distributed with two degrees of freedom.
The Catalina V-band light curves are rebinned with minimal 1-day time-lag, and the average number of updated light curve 
data is $\sim$70. When the GLSdeDRW fitting power $p = \chi^2_0 - \chi^2(f) \gtrsim$ 8,  the signal is significant ($FAP$ $\lesssim$ 1\%).

The ACF and CCF analysis are applied to avoid the false detections from {\it GLSdeDRW}. The DRW-like 'noise' in quasar light curve would generate an exponential decaying exp(-t/$\tau$) correlation function in the CCF analysis. To exclude the DRW-like 'noise' in quasar light curves, we use Equation (1) to diagonalize the covariance matrix for each quasar, then linearly transform this to generate a 'white noise' light curve. The standard ACF test is applied on these modified light curves. For quasars with period $T_0'$ and covariance time scale $\tau_0'$ found by {\it GLSdeDRW}, if $T_0'$ $>>$ $\tau_0'$, the DRW-like 'noise' in CCF period searching is negligible, and we can directly apply CCF analysis on their original light curves.
We select the Z-transformed discrete correlation function code 
\citep[ZDCF,][]{Alexander97} for the following ACF and CCF analysis. 
The ZDCF code is demonstrated in a small sample of sparsely sampled light curves.  
The significance level of ACF and CCF analysis is set as the two-side critical value \citep[][originally from Anderson 1941]{Salas80}:
\begin{equation}
r_{\Delta t}(P=99\%) = \frac{-1 \pm 2.326\sqrt{N'-2}}{N'-1},
\end{equation}
here $N'$ is the sample size in each time-lag $\Delta t$ bin, and we require $N'$ $\geq$ 10 in the correlation analysis.

We apply the {\it GLSdeDRW} and ACF tests on the CRTS $V$-band light curves of the 347 low-$z$ bright quasars. 
The SMBHB candidates are selected when the period $T$ is significant ($P \gtrsim$\,99\%) in both {\it GLSdeDRW} and ACF tests. 
Three quasars are selected as the SMBHB candidates. Only one of them, 
SDSS\,J015910.058+010514.53 (hereafter SDSS\,J0159+0105), has no FIRST radio detection.  SDSS\,J0159+0105 also shows 
variable broad balmer line profiles in its two SDSS spectroscopic observations. 
The detailed observational properties of SDSS\,J0159+0105 are summarized and analyzed in the following two sections.

\section{Observational Properties of SDSS J0159+0105}

SDSS J0159+0105 is a broad line quasar at $z=$ 0.217 in SDSS Stripe 82 field. It has a companion-galaxy at the same redshift but 6\arcsec\ north.
The CRTS V-band light curve of SDSS J0159+0105 is plotted in Fig. \ref{ppn129lc}, as well as
the light curve analysis with our code {\it GLSdeDRW} and ACF. The detailed time-series analysis is presented in Sec. \ref{Sec:LC}.

\begin{figure*}
\begin{center}
\includegraphics[angle=270,width=\linewidth]{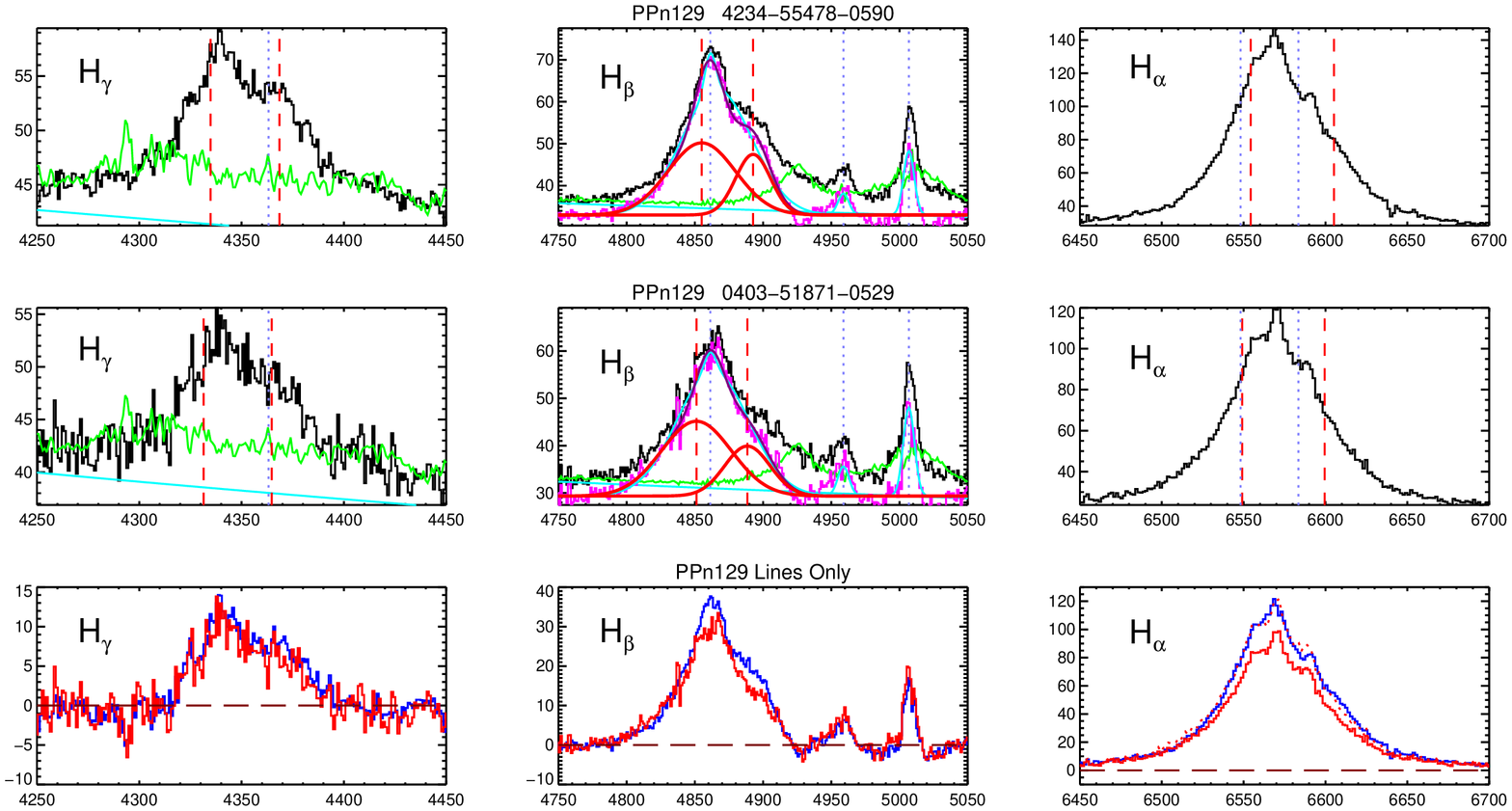}  
\caption{Two-Gaussian fitting results to the broad H$\beta$ lines of the quasar SDSS J0159+0105. As a comparison, we also plot the H$\alpha$(right column) and H$\gamma$ (left column).  The SDSS spectra in two epochs are presented in the top and middle panels. In the bottom panel, the continuum and Fe{\sc II} template subtracted spectra of the two epoch are presented. In each stamps we mark the narrow emission lines in light-blue vertical dotted lines. The center wavelength positions of two gaussian fitting to the broad H$\beta$ are marked in red vertical dashed lines, and scaled to the stamps showing H$\alpha$ and H$\gamma$.  }
\label{hb2gfit}
\end{center}
\end{figure*}

SDSS\,J0159+0105 has two spectroscopic observations taken by SDSS at MJD = 51871 and by BOSS at MJD = 55478, with median $r$-band S/N of 24 and 48 per pixel, respectively. 
We follow \citet[][and reference there in]{Shen13} to fit the profile of the broad H$\beta$  line. In brief, a power-law continuum and an iron emission line template \citep{BorosonGreen92} 
are fitted and subtracted, then the regions of narrow lines are fitted with narrow gaussian lines, and the asymmetric broad  H$\beta$ line is fitted with two gaussians. 
The fitting results are presented in Tab. \ref{2spec}, and the H$\beta$ + [O\,{\sc iii}] emission lines from the two epochs are presented in Fig. \ref{hb2gfit}.
Obviously, the [O\,{\sc iii}] lines did not change in the two epochs, while the broad H$\beta$ line showed a larger but narrower red bump with higher red-ward offset-velocity in the second epoch.  
The implication of the broad H$\beta$ is presented in Sec. \ref{Sec:Hb}.

SDSS\,J0159+0105 has no FIRST radio detection (see the FIRST stamp in Fig. \ref{sed}), and the RMS noise is 0.135 mJy\footnote{{\it FIRST} archive link: http://sundog.stsci.edu}.
It is excluded as a radio-loud AGN from the loudness limit \citep{Sikora07}.  Compared to the typical SED of radio-quiet quasars \citep{Shang11}, the SED of SDSS\,J0159+0105
is matched in the radio, optical, and X-ray bands, but shows extra radiation in the infrared and UV bands. The SED is presented in Tab. \ref{photo} and plotted Fig. \ref{sed}. 
 We briefly discuss the SED of SDSS\,J0159+0105 in Sec. \ref{Sec:SED}.

\begin{figure}
\begin{center}
\includegraphics[angle=270,width=\linewidth]{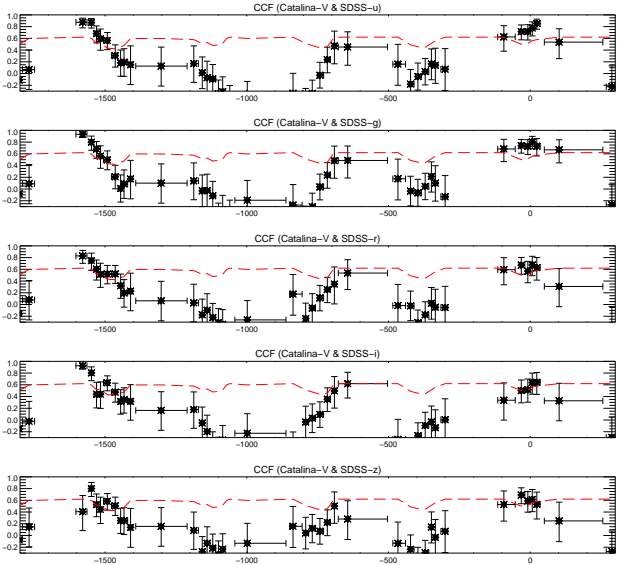}
\caption{The cross-correlation analysis (CCF) between CRTS V-band light curve and SDSS $ugriz$ band light curves. The red dashed lines show the $P>$ 99\% significance level. Obviously, the signals at $\Delta T$ $\sim$ -1579 days are 
significant in all the CCF analysis. 
}
\label{ccf}
\end{center}
\end{figure}

\section{Physical Interpretation}

In this section, we examine the above three observational properties of  SDSS\,J0159+0105. We show that
 the periodic light curve points to the existence of a SMBHB in this quasar. The profile change of
broad H$\beta$ line and SED can be linked to the existence of a SMBHB in this quasar, however, there are alternative explanations.

\subsection{Evidence 1: Periodic Light Curve }
\label{Sec:LC}

Our {\it GLSdeDRW} method resolves two (three) periods above $99\%$ ($95\%$) confidence level in the Catalina $V$-band light curve with covariance time scale $\tau_0$ $\sim$ 210 day and variability scale $\hat \sigma$ $\sim$ 0.005. Only the 
sinusoidal signal, which can be fitted as $V$ =  0.1 $\times$ sin (2$\pi$(t - 53413.09 )/741) + 16.59 (see Fig.~\ref{ppn129lc}-a), is also confirmed by the additional ACF analysis ($P$\,$>$\,99\%). We found no similar periodicity in the bright sources close to SDSS\,J0159+0105 ($<$\,1-arcmin), suggesting that the sinusoidal signal we found is real.
The periodic signals at $T\sim$ 200--300 day shown in the {\it GLSdeDRW} test are likely false signals as they do not show up in the ACF analysis. 
Interestingly, from the ACF analysis alone we also found a second periodic signal at T = 1497$^{+20}_{-62}$ day, which has much stronger significance, and 
is approximately twice the period of the sinusoidal signal derived from the {\it GLSdeDRW} method. This second ACF signal is insignificant in the {\it GLSdeDRW} test, the reason is that 
the {\it GLSdeDRW} test is designed to search for the sinusoidal signals, while ACF is sensitive to any periodic signals.

Although the poor cadence of SDSS light curves prevents the search of the periodic signals, we can cross-correlate the CRTS $V$-band light curve and 
the SDSS five-band light curves to find a possible periodic delay (the MJD baselines of SDSS and CRTS light curves are 52170--54433 and  53627--56591, respectively). After doing this cross correlation, 
we found a strong and significant ($P>$ 99\%) correlation signal at $\Delta T$ = -1579$^{-24}_{+15}$ day between the CRTS $V$-band light curve and each of the SDSS $ugriz$ band light curves. This strong correlation 
confirms that the periodic signal at $T$ = 1497$^{+20}_{-62}$ day from the ACF test is robust (see Fig. \ref{ccf}).

Since SDSS\,J0159+0105 is a radio-quiet quasar, we can exclude the jet-procession hypothesis and directly link the periodicity to the existence of a SMBHB. 
Recent simulations \citep[e.g.,][ and references there in]{DOrazio13,DOrazio15a} predicted that the accretion rate of a SMBHB 
has two significant periods that are at the $2:1$ ratio. For example, \citet{DOrazio15a} showed that
when $0.05<q<0.3$, the only two prominent periods are $t_{orb}$ and $0.5t_{orb}$, which are directly linked to the orbital motion of the binary.
When $0.3<q<0.8$, although an additional, more prominent period would appear, which is about $(3-8)t_{orb}$ and related to the lumpy structures at the
verge of the circumbinary disk, the 
$t_{orb}$ and $0.5t_{orb}$ periods remain significant. Only when $q\simeq1$ does the period of $t_{orb}$ disappear.

For SDSS\,J0159+0105, since we also found two periods that have a ratio of $2:1$, namely $1497$ and $741$ days, we identify them with the
characteristic $2:1$ periods proposed by  \citet{DOrazio15a}. This consideration leads to a binary orbital period of $t_{orb} = 1497$ days (or 3.3 year in the rest-frame). The $2:1$ period ratio also indicates the unlikely of an equal-mass binary inside SDSS\,J0159+0105.

\subsection{Broad H$\beta$ Line and Its Profile Variability: the Second Evidence?}
\label{Sec:Hb}

Broad lines in AGNs are powerful probes of the physical parameters of the central SMBHs
\citep[e.g., see the review of BLR by][]{Peterson06}. 
It is well established that there is a tight correlation between the BLR radius and continuum luminosity
 derived from reverberation mapping of local AGNs \citep[e.g.,][]{ VestergaardPeterson06,Bentz09}.
 We use this correlation to estimate the virial mass of the black-hole in our system. 
  Given the size of the BLR, 0.04pc or 50$^{+30}_{-20}$ light-day, inferred from the $R-L$ correlation, we find that the virial mass is 10$^{8.11\pm0.11}$ M$_\odot$ inside SDSS\,J0159+0105 (from BOSS spectrum). 
The total mass gives a SMBHB separation of $d$ = (0.01pc)$M_{8,tot}^{1/3}$ (T/3.3yr)$^{2/3}$ $\sim$ 0.013pc  (15 light-day). 
This centi-parsec separation, which is comparable to the size of the BLR, excludes the existence of two distinctive BLRs surrounding each SMBH in the binary system, and supports a single circumbinary BLR surrounding this SMBHB.

The variation of the broad H$\beta$ line profile can be explained in the single circumbinary BLR model. 
The profile of the broad H$\beta$ line can be resolved into two gaussian bumps in both spectra (see Tab \ref{2spec}): a strong blue bump located in the center, 
and a small red bump shifted red-ward. While the relative velocity varies, the flux ratio of the two bumps is nearly stationary, and 
the line widths are consistent (within 1--2$\sigma$) with being a constant. 
In the single circumbinary BLR model, the BLR clouds attaining the photoionization equilibrium suffer 
the ionizing radiation from the SMBHB, $U$ = $Q_{\rm ion, P}(H)/(4\pi r_P^2cn_e)$ + $Q_{\rm ion, S}(H)/(4\pi r_S^2cn_e)$. 
Here $Q_{\rm ion, P}(H)$ and $Q_{\rm ion, S}(H)$ are H-ionizing photons from the primary and secondary SMBHs, and $r_P$ and $r_S$ are the distances of the two BHs to the 
specific Keplerian BLR clouds, respectively.   
When the SMBHB mass ratio $q$ is small and the accretion is dominated by the secondary SMBH, the periodic off-nuclei radiation from the secondary SMBH will generate the asymmetric profiles of broad lines. A detailed description of the connection is left to future work.

The shifted red bump can also be explained as the inflow stream into the secondary SMBH. Shined by the UV continuum
from the mini-disks, the inflow stream should have a varied viewing angle led by the rotation of the SMBHB.
This will introduce the coincidence of inverse relation between offset-velocity and line-width of the inflow stream: the larger the offset-velocity, the narrower the width for the small bump.  
We do find this anti correlation from the two-epoch spectroscopic observations, however, the limited data prevent us from constraining the inflow hypothesis. 
In the inflow model, we would expect the offset-velocity, the line-width, and their anti correlation relation of the small bump vary periodically. 
This periodic variable spectral profile will also show up under the BLR clouds model, but with no inverse correlation. 
Another difference between the two model is that under the inflow stream case, we would expect an extreme condition that appears in a short time and periodically. When the inner part of the inflow is along the line-of-sight direction, the red bump would have much larger offset-velocity, but much narrower line-width. The detailed interpretation is beyond the topic of this work, and we would like to explore it with future spectroscopic monitoring data.

We notice that, although the red-bump in broad H$\beta$ is significant, it is not very significant in broad H$\alpha$ (BOSS spectrum). 
When the ionization parameter increases for a constant plasma, a decrease of $F(H\alpha)$/$F(H\beta)$ intensity ratio
is expected \citep{Wills85}. This is due to the increase of the excitation state of the ionized gas: the temperature of the ionized zone being higher, the population of the 
upper levels with respect to the lower ones increases. However, both the BLR clouds and inflow streams can explain this. 
In the BLR clouds model, the inner clouds in BLR should be more compact and have higher temperature than the outer clouds.  
Since the secondary SMBH is closer to the inner clouds, there would be a larger covering factor for the inner compact 
clouds to the secondary SMBH than that to the primary SMBH, which lead to a decreased $F(H\alpha)$/$F(H\beta)$ intensity ratio of the red-bump.
For the inflow stream model, the larger the offset-velocity of the red bump is, the closer the stream is to the secondary SMBH. The inner inflow streams should have higher temperature thus higher ionization parameter, which also  lead to a decreased $F(H\alpha)$/$F(H\beta)$ intensity ratio of the red-bump.

The varied broad H$\beta$ line profile of SDSS\,J0159+0105 provides an ideal probe to investigate the dynamics of gas clouds/streams
around the SMBHB. However, unlike the previous evidence for the existence of SMBHB, the broad H$\beta$ and its profile variability from only two-epoch spectroscopic observations
can not be used as evidence of SMBHB, until we find the periodic shifts of the red bump from further spectroscopic observations.

\subsection{SED with Extra Bright UV Radiation: the Third Evidence?}
\label{Sec:SED}

\begin{figure*}
\begin{center}
\includegraphics[angle=0,width=\linewidth]{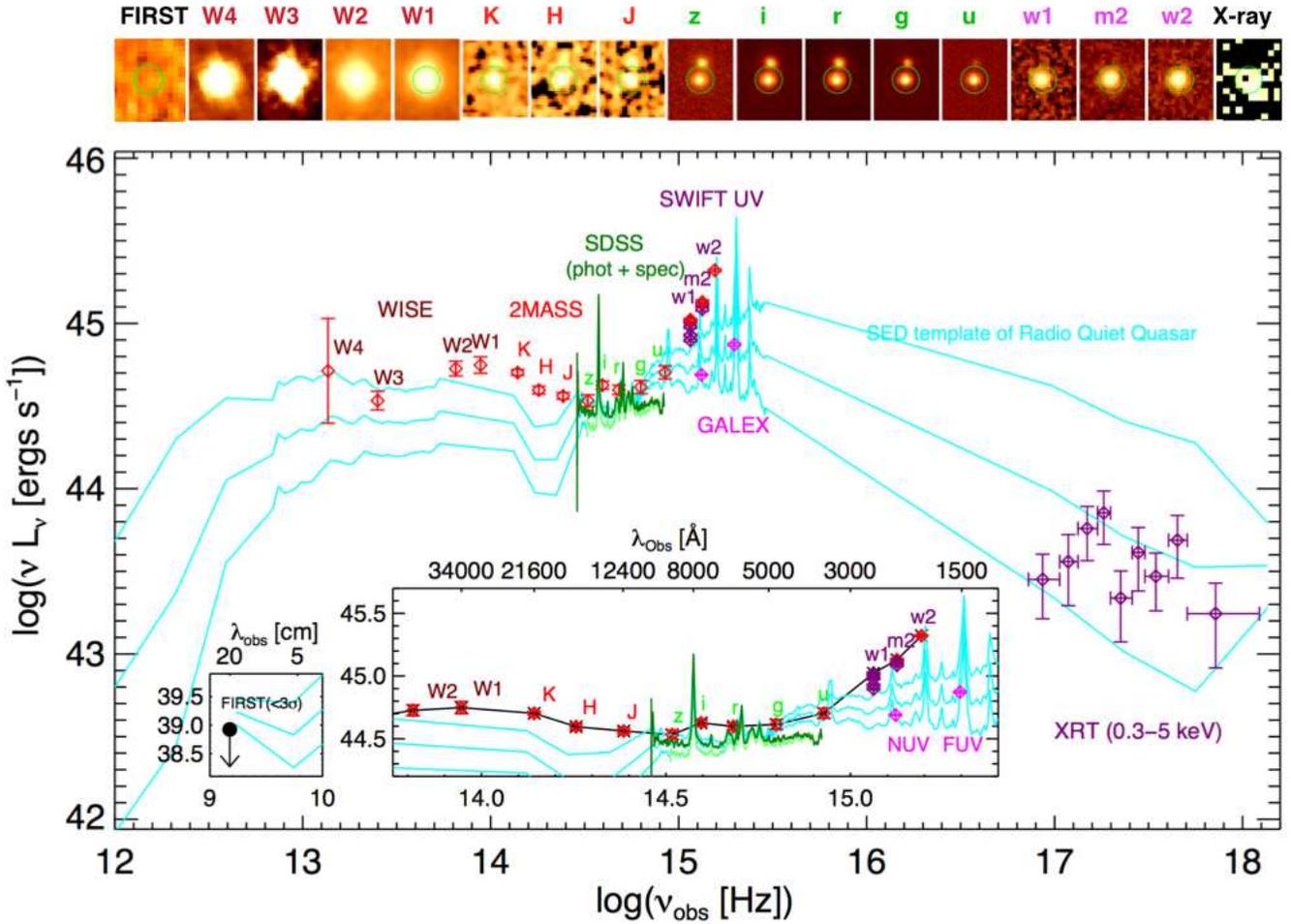}  
\caption{The SED of  SDSS\,J0159+0105, measured from FIRST, \textit{WISE}, 2MASS, SDSS, {\it GALEX} UV, {\it SWIFT} UV and XRT. The SED values are presented in Tab. \ref{photo}. 
On the top panel the stamps at different energy bands are presented. The size of the stamps in each band is 25\arcsec$\times$ 30\arcsec\, 
with a $r$=5\arcsec\ circular marking the position of SDSS\,J0159+0105. 
In the main panel, the SDSS and BOSS optical spectra of SDSS\,J0159+0105 are plotted in light-green and dark-green colors. The cyan curves represent the composite SED of
radio-quiet quasars \citep[median plus $\pm$1$\sigma$ SED,][]{Shang11}, including the {\it HST} UV composites \citep{Telfer02} and the SDSS composite \citep{VandenBerk01}.  
Notably, the UV and IR radiation of SDSS\,J0159+0105 are significantly brighter (3.2 and 2.0 times in the {\it SWIFT} $w2$ and 2MASS K bands, respectively) than the SED template of radio-quiet quasars normalized in the SDSS-$g$ band. However, there is no simultaneous optical and NIR measurements in the epoch of {\it SWIFT} UV burst (MJD = 57082--57291). 
Thus the UV variability could be related to the binary nature of the SDSS\,J0159+0105, but we can't exclude the UV flare from normal AGN. 
}
\label{sed}
\end{center}
\end{figure*}

The schematic diagram of SMBHB is comprised of two mini-disks around each SMBH connected with streams, a cavity or gap opened by the rotation of the secondary SMBH, and 
a circumbinary disk embedding the innermost SMBHB, mini-disks and cavity system \citep[e.g.,][]{ArtymowiczLubow94,Cuadra09}. Compared to the standard AGN disk, the SMBHB structure will lead to distinctive features in the SED, 
such as a deficit in the UV-optical spectrum of Mrk 231 reported by \citet{Yan15}, or an enhanced UV radiation predicted by Lodato et al. (2009) and Farris et al. (2015).
Here in the SED of SDSS\,J0159+0105, we find similar flux levels to the typical SEDs of radio-quiet quasars \citep[][normalized in the SDSS-$g$ band]{Shang11} at radio, optical and X-ray
frequencies, but 3.2 and 2.0 
times brighter in the {\it SWIFT} $w2$ and 2MASS $K$ bands (see Fig. \ref{sed}, notice that the extra radiation in UV is more significant than that in IR). Although there are strong emission lines covered by the {\it SWIFT} UV bands (i.e., N\,{\sc iv}], C\,{\sc iv} and Fe broad features in {\it SWIFT} uvw2, Al\,{\sc iii} and C\,{\sc iii}] in {\it SWIFT} uvm2, and Fe\,{\sc ii} in {\it SWIFT} uvw1), the observed equivalent widths contributed by emission lines of radio-quiet quasars \citep[see Table 2 of][]{Telfer02} in the three UV bands are 112\AA, 27\AA, and 1\AA\ in {\it SWIFT} uvw2, uvm2, and uvw1 bands, respectively.  The effective band widths of {\it SWIFT} uvw2, uvm2, and uvw1 filters are 610\AA, 519\AA, and 795\AA, respectively. So the contribution of the emission lines to the UV flux are only 0.07dex, 0.02dex and 0.00dex in {\it SWIFT} uvw2, uvm2, and uvw1 bands, respectively. The emission-line subtracted UV radiation is 2.7 times brighter than the template. 

The large variability of UV radiation captured by {\it GALEX} and {\it SWIFT}  agrees with the prediction by \citet{Tanaka13}, who has interpreted that the UV light curves of SMBHB in one period can 
dim by 2 orders of magnitude. However, we notice that there is no simultaneous optical and NIR measurements in the epoch of {\it SWIFT} UV burst (MJD = 57082--57291). Thus the UV variability could be related to the binary nature of the object, but we can't exclude the UV flare from normal AGN. Moreover, the extra infrared radiation is within $\pm$2$\sigma$ of the template. 
So the current SED data is not a proof of SMBHB in the quasar SDSS\,J0159+0105.

\subsection{The Unified Picture of the SMBHB inside SDSS\,J0159+0105}

In summary, we have one observational feature, the periodic light curves, supporting the existence of a SMBHB inside SDSS\,J0159+0105. 
Since SDSS\,J0159+0105 is a radio-quiet quasar,  we can easily exclude the jet contribution to the periodic light curves and directly link the periods to the rotation of SMBHB. 
The varied broad H$\beta$ profiles from two spectroscopic observations may imply a single circumbinary BLR embedding the SMBHB, or inflow streams connecting the SMBHB and 
the circumbinary disk, or both. This can not be taken as evidence of SMBHB unless we find the periodically varied profile of the broad H$\beta$ line in the future.  
  The varied UV SEDs may imply the existence of SMBHB, however, we can't exclude the normal AGN flare hypothesis.

The physical properties of the SMBHB inside SDSS\,J0159+0105 are probed from these observational features. 
The SMBHB has an orbital period of $t_{orb}\sim$ 1497 day (from time-series analysis), and a total mass of $\sim$ 1.3$\times$10$^8$ M$_\odot$ (from virial estimation). 
The binary separation is $d$ $\sim$ 0.013~pc (15~light-days). 
The circumbinary disk in the SMBHB system can be approximated as a standard AGN disk truncated by a cavity \citep{Tanaka13}. The cavity radius (the inner radius of the circumbinary disk) is $\approx 2a$ \citep{ArtymowiczLubow94}, where $a$ is the semi major axis of the binary and 2$a$ is the Lindblad resonance radius.
Here the radius of the cavity is 2$a$ = 2$d$ $\sim$ 0.026~pc. 
There is a single circumbinary BLR in the BBH system with an average radius of $\sim$0.04 pc($\pm$0.3~dex uncertainties, from $L$-$R$ relations of local AGN). Interestingly, the inner radius of the circumbinary disk and the BLR size, measured from two independent methods, are approximately similar. 

To acquire a stable cavity and circumbinary disk in the SMBHB, it is required $a$ $\gtrsim$ $a_{dec}$ $\sim$ 100$GM/c^2$ \citep{MilosavljevicPhinney05,Haiman09,TanakaMenou10}. 
Here $a_{dec}$  is a `decoupled' separation when the close enough binary starts to shrink faster than the local viscous timescale, 
i.e., the energy release by the gravitational wave is faster than that by the disk. 
During the lifetime of the binary, the surface density of the circumbinary disk peaks at a radius $\sim$max(2$a$, 2$a_{dec}$) \citep{Tanaka13}.
In our case, the system is quite stable because of $a$ $\sim$ $d$ $\gtrsim$ 10$a_{dec}$. In the current stage, the binary could have been shrinking mainly because of gravitational wave emission,
 thus the system is in a gravitationally bound regime. 

\citet{DOrazio15b} proposed that the relativistic beaming of the radiation from the mini-disk around the secondary SMBH is sufficient to explain the periodicity in PG 1302-102. 
In their case, the total mass of the SMBHB in PG 1302-102 should be above 10$^{9}$ M$_\odot$ and the secondary SMBH should dominate the radiation. Although we have nearly
the same period, the total mass of the SMBHB in SDSS\,J0159+0105 is only 10$^{8}$ M$_\odot$. The relativistic doppler boosting calculated by \citet{DOrazio15b} is $\Delta F_v/F_v$ = $\pm$(3-$\alpha$)$(v {\rm cos}\phi$/c)sin$i$. Here $i$ and $\phi$ are the inclination and phase of the orbit, and $\alpha$ is the frequency index ($F_\nu \propto \nu^\alpha$, see Tab. \ref{2spec}). The rotation velocity of the secondary SMBH $v$ is determined as $v$= ($\frac{2\pi}{1+q}$)($\frac{GM_{\rm SMBHB}}{4\pi^2T_{r}}$)$^{1/3}$$\sim$ 8000 km s$^{-1}$.  
In the virial BLR assumption, we have sin$i$ = max([$\Delta V_1, \Delta V_2$])/$\sqrt{2GM_{\rm SMBHB}/R_{\rm BLR}}$ $\sim$  0.4.
So the maximum boosting factor is $\Delta F_v/F_v$ = $\pm$(3-$\alpha$)($v$/c)sin$i$ $\sim$ 0.03, which is less than the amplitude from our $V$ band observations of $\Delta F_v/F_v$ $\sim$ 0.1. 
We conclude that the relativistic beaming effect should not dominate the periodic signal in SDSS\,J0159+0105. The time-dependent accretion (caused by the orbital motion of the binary) displays the 0.5$t_{orb}$ and $t_{orb}$ periods in the light curves of SDSS\,J0159+0105.

\section{Conclusions}

In this work, we report a strong SMBHB candidate in the radio-quiet quasar SDSS\,J0159+0105 at $z=0.217$, which has periodic optical light curves, broad balmer lines with variable intensities and profiles, and an unusual SED with extra bright UV radiation. The existence of the SMBHB can be verified by the first observational property. The combination of the observational properties reveals a 
1.3$\times$10$^8$ M$_\odot$ SMBHB with a binary separation of $d$ $\sim$ 0.013~pc (15 light-day). The SMBHB has a circumbinary disk with an inner radius of $\sim$0.026pc. There is a circumbinary BLR surrounding the SMBHB, and the average radius of the BLR is $\sim$0.04pc (50 light-day, with $\pm$0.3 dex uncertainties). The time-dependent accretion (caused by the orbital motion of the binary)  determines the periodic light curve.

Till now, the most significant SMBHB candidates selected from time-series analysis includes PG 1302-102 \citep{Graham15a}, OJ 287 \citep{Valtonen08}, and PSO J334.2028+01.4075 \citep{Liu15}. 
All of their host quasars are radio-loud, thus the periodic variability might be caused by the jet precession. 
Recently, \citet{Graham15b} reported a systematic search of 111 SMBHB candidates from CRTS data. We notice that most of which are hosted by radio-quiet quasars, but our candidate SDSS\,J0159+0105 is not in their catalog. Only one candidate UM 234 in their catalog is covered in our search. We find that UM 234 has $P>99$\% period in the ACF test, while the signal is at $P\sim$ 95\% with our {\it GLSdeDRW} test. As a comparison, we apply the {\it GLSdeDRW} test with Graham's most significant SMBHB candidate PG 1302-102 \citep{Graham15a}. {\it GLSdeDRW} yields the same signal as they reported at a significance level of $P$ $>$99.999\% from our {\it GLSdeDRW} test. 
We would like to start our systematic search for close SMBHBs with the all sky quasar light curves (Zheng \& Butler, in prep), and prepare for the upcoming Large Synoptic Survey Telescope \citep[LSST,][]{Ivezic08}.

\section*{Acknowledgements}

We are grateful to Matthew Graham and Zoltan Haiman for valuable comments, and to the anonymous referee for his/her helpful comments and suggestions. ZYZ gratefully acknowledges support from CAS-CONICYT postdoc fellowship.  This work has been developed during the stay of ZYZ as SESE Exploration postdoctoral fellow at Arizona State University.  XC and JC are supported by CONICYT-Chile through Anillo (ACT1101) and the ``VRI concurso estad\'{i}as en el extranjero'' of PUC. JC acknowledges support from CONICYT-Chile through FONDECYT (1141175), Basal (PFB0609) and Anillo (ACT1101) grants.

\clearpage

 \begin{table*}
\caption{ Photometric informations of SDSS\,J0159+0105. }
\begin{center}
\begin{tabular}{cccccc}\hline\hline
 RA &  01:59:10.058    & Dec &  +01:05:14.53 & & redshift =  0.217, E(B-V) = 0.0232  \\ \hline\hline
Band & Instrument + filter  & \multicolumn{2}{c}{Photometry} &  MJD   &  \\   \hline
\textbf{X-ray}   &             & [keV]  & [ erg cm$^{-2}$ s$^{-1}$]      \\
&  SWIFT-XRT & 0.3-10 keV & 7.4$^{+1.9}_{-1.9}$ $\times$10$^{-13}$   & 57082 &  $\Gamma$ = 2.6$^{+0.8}_{-0.6}$ \& N$_H$ = 3.0$^{+20.4}_{-3.0}\times 10^{20}$  cm$^{-2}$  \\
      &       & $\lambda_{center}$ [\AA] & Mag$_{AB}$   &    &  \\
\textbf{UV}        &  &  & &   \\
 & GALEX-FUV & 1516 & 17.81$\pm$0.02 & 53697 &   \\
     &  GALEX-NUV & 2267 & 17.83$\pm$0.01 &  53697  &  \\
  &SWIFT-uvw2 &  1928  & 16.43$\pm$0.02  &  56813 & \\
  &SWIFT-uvm2 & 2246 & 16.85$\pm$ 0.02 & 57082 &  \\
  &---   & ---   & 16.78$\pm$ 0.02 & 57286 & \\
  &SWIFT-uvw1 & 2600 & 16.85$\pm$0.02 & 57083 & \\
  & ---   & ---   & 16.89$\pm$0.02 & 57083 & \\
  &  ---   & ---   & 16.96$\pm$0.02 & 57179 & \\
  &   ---   & ---   & 17.16$\pm$0.02 & 57183 & \\
  &    ---   & ---   & 17.08 $\pm$0.02 & 57291 & \\
 \textbf{Optical} &         &  &  &  &  \\
 &SDSS-u &  3543 & 17.31$\pm$0.10 &  52170--54433  & \\
 &SDSS-g &  4770 & 17.21$\pm$0.10 &  52170--54433  & \\
 &SDSS-r &  6231 & 16.96$\pm$0.10 &  52170--54433  & \\
 &SDSS-i &  7625 & 16.67$\pm$0.10 &  52170--54433  & \\
 &SDSS-z &  9134 & 16.71$\pm$0.10 &  52170--54433  & \\
 & CRTS-V & 5300 & 16.59$\pm$0.06 & 53627--56591 & \\
\textbf{Infrared} &          &  &  &  &  \\
 & 2MASS-J & 12350 &  16.31$\pm$0.05 &  51784 &  \\
 & 2MASS-H & 16620  & 15.90$\pm$0.06 & 51784 &  \\
  &2MASS-K & 21590 & 15.35$\pm$0.05 & 51784 &  \\
 & WISE-W1  & 34000 & 14.74$\pm$0.12 & 55210--55576 &  \\
  & WISE-W2  & 46000  & 14.47$\pm$0.11 & 55210--55576 &  \\
 & WISE-W3  & 120000 & 13.91$\pm$0.14 & 55210--55576 &  \\
 & WISE-W4  & 220000 & 12.80$\pm$0.79 & 55210--55576 &  \\
 \textbf{Radio}      &  & $\lambda$ [cm] & $F_{<3\sigma}$ [mJy] &  \\
 & FIRST  &  20  & $<$0.41    & 50006 &   \\
\hline
\end{tabular}
\tablecomments{1). Although the companion galaxy is unresolved in the {\it WISE} images, we can get the contribution of the companion galaxy from the well resolved {\it 2MASS} images. 
In $JHK$ bands, the flux density ratios of SDSS\,J0159+0105 and its companion galaxy  are 3:1, 4:1 and 4.6:1, respectively.
So the contribution from the unresolved companion galaxy in {\it WISE} images should be less than 25\%. \\
2). The galactic dust extinction is E(B-V) = 0.0232 \citep{SchlaflyFinkbeiner11}, which is very small and gives extinction corrections $<$ 0.1 dex for the above UV and optical bands. So in this paper we didn't apply the extinction correction in the SED.  }
\end{center}
\label{photo}
\end{table*}


\begin{table*}
\caption{ Spectral fitting results of SDSS\,J0159+0105. }
\begin{center}
\begin{tabular}{ccccc}\hline\hline
  \multicolumn{5}{c}{H$\beta$ Fitting Results with different Instruments and MJD}  \\
  Component & Parameter & Unit.  & SDSS--51871 & BOSS-55478 \\ \hline
   Continuum         & log(L$_{5100}\lambda_{5100}$)  & [erg s$^{-1}$] & 44.39$\pm$0.03   & 44.44 $\pm$0.02   \\   
                                &  $\Gamma_{5100}$  & --            & 1.88$\pm$0.02  & 1.57$\pm$0.01   \\
Blue-Bump  &  $\Delta V_1$   &  [km s$^{-1}$]  & -625$\pm$537  & -390$\pm$321   \\ 
			& $FWHM_1$ &  [km s$^{-1}$]  & 1580$\pm$241& 1530$\pm$123   \\ 
			& $F_1$ &  [10$^{-17}$ erg s$^{-1}$ cm$^{-2}$] & 1014$\pm$319  &  1068$\pm$217   \\
  Red-Bump   & $\Delta V_2$  &  [km s$^{-1}$]  &  1677$\pm$210  &  1936$\pm$93 \\
                      & $FWHM_2$  &  [km s$^{-1}$]     &   1024$\pm$123   &   790$\pm$80 \\ 
                  & $F_2$  &  [10$^{-17}$ erg s$^{-1}$ cm$^{-2}$]  & 440$\pm$269   & 464$\pm$155 \\
\hline
\end{tabular}
\tablecomments{$F_\lambda \propto \lambda^\Gamma$, and $F_\nu \propto \nu^\alpha$, here $\alpha$ = 2-$\Gamma$}
\end{center}
\label{2spec}
\end{table*}


\label{lastpage}

\end{document}